\documentclass[9pt,twocolumn,twoside]{osajnl}

\journal{optica}

\setboolean{shortarticle}{true}

\usepackage{xcolor}
\usepackage{siunitx}
\usepackage{graphicx}
\usepackage{amsmath}                                                          
\usepackage{textcomp}
\usepackage{listings}
\usepackage{courier}
\usepackage{booktabs}
\usepackage{nicefrac, xfrac}

\graphicspath{{img/}}

\usepackage{todonotes}
\presetkeys{todonotes}{inline,inlinewidth=4cm}{}

\begin{document}

\title{Observation of Noise Suppression during High-Efficiency Wavelength Doubling of Intense Quasi-Monochromatic Laser Light}

\author[1,*]{Julian Gurs}
\author[1]{Mikhail Korobko}
\author[2]{Christian Darsow-Fromm}
\author[3,4]{Sebastian Steinlechner}
\author[1]{Roman Schnabel}

\affil[1]{Institut für Quantenphysik und Zentrum für Optische Quantentechnologien, Universität Hamburg, Luruper Chaussee 149, 22761 Hamburg, Germany}
\affil[2]{Institut für Experimentalphysik, Universität Hamburg , Luruper Chaussee 149, 22761 Hamburg, Germany}
\affil[3]{Faculty of Science and Engineering, Maastricht University, Duboisdomein 30, 6229 GT Maastricht, The Netherlands}
\affil[4]{Nikhef, Science Park 105, 1098 XG Amsterdam, The Netherlands}
\affil[*]{jgurs@physnet.uni-hamburg.de}

\begin{abstract}
Ultra-stable, quasi-monochromatic laser light forms the basis for high-precision interferometric measurements, e.g.~for observing gravitational waves (GWs) and for time keeping with optical clocks. Optical frequency conversion enables access to wavelengths at which optical materials have the lowest absorption and the lowest mechanical loss. Here we report a 25\,\% reduction in  relative intensity noise (of technical origin) when converting 1064\,nm to 2128\,nm for powers far above parametric oscillation threshold. The new wavelength has high potential for improving GW detection and other ultra-high-precision experiments as well.
Our results provide a better understanding of the dynamics of nonlinear optical processes and have great potential for the stabilisation of laser sources in optical sensing and metrology.
\end{abstract}

\maketitle

\section{Introduction}
Modern ultra-high precision experiments, such as optical atomic clocks and gravitational wave detectors, require an unprecedented level of stability in the laser light used~\cite{Ludlow2015,GW150914}. Intrinsically stable laser devices with typical wavelengths at 1064\,nm or 1550\,nm are employed~\cite{GW150914,Kessler2012} in combination with intricate stabilization schemes for the light's intensity (or `amplitude', `power') and frequency (or `phase')~\cite{Hall1978,Robertson1986,Salomon1988,Kwee2012}. 
The transmission of light through filter resonators passively reduces the noise at frequencies above the line width of the resonator. At lower frequencies, part of the light must be measured and the remaining part actively stabilized, which has the disadvantage of introducing additional photon shot noise and electronic control noise. Ultra-high precision experiments are also limited by thermal energy and optical materials with too high mechanical loss~\cite{Matei2017,Robinson2019,Robinson2021} and might benefit from silicon-based optics combined with a slightly increased laser wavelength in the short-wave infrared regime around \SI{2}{\micro\metre}~\cite{steinlechnerSiliconBasedOpticalMirror2018a}. Additional benefits of longer wavelengths are the reduction of Rayleigh scattering and wavefront distortion due to imperfect surface polish.\\
If intrinsically stable laser devices are not available at the desired wavelength, frequency conversion via nonlinear optics is an option. Recently, we proposed degenerate optical parametric oscillation far above threshold for wavelength doubling of the widely available ultra-stable laser light at 1064\,nm~\cite{Darsow-Fromm2020} and demonstrated the transfer of the ultra-low-noise property of a 25\,W beam at 1064\,nm to a 16\,W beam at 2128\,nm~\cite{Gurs2023}.
Nonlinear optics was previously used for the noise suppression of intense quasi-monochromatic light~\cite{Khalaidovski2009,Thuering2011}. In these works, the cascaded optical Kerr effect was used, which does not provide frequency conversion, which is what our current research and development aims for.\\
Here, we report the first observation of intensity noise suppression via degenerate optical parametric oscillation (DOPO) far above its lasing threshold. 155\,mW of down-converted light at 2128 nm showed a 25.1(18)\% lower (technical) relative intensity noise (RIN) than the initial already very stable 275\,mW at 1064\,nm. The noise suppression is achieved for sideband frequencies well inside the linewidth of our conversion resonator of several tens of MHz. The noise reduction is passive and does not require measurement of any light, which avoids additional photon shot noise and other control noise. We present a comparison of the RIN noise spectra of the two wavelengths for three different pump powers focusing on a low-frequency band with a rather flat spectrum and low detector dark noise between \SI{90}{\kilo\hertz} to \SI{100}{\kilo\hertz}. Our result is in very good agreement with our theoretical model, which suggests a potential maximal suppression factor of 2 at infinite pump powers, consistent with earlier theoretical studies~\cite{Walls1990,Wiseman1991}.\\
We note that {\it degenerate} optical parametric oscillation operated far above oscillation threshold has been rarely used in the past, while {\it non}-degenerate OPO is a standard approach to convert laser radiation at frequency $\omega_0$ to signal $\omega_\text{sig}$ and idler $\omega_\text{id}$ fields at frequencies obeying energy conservation ($\omega_\text{sig}+\omega_\text{id} = \omega_0$)~\cite{Harris1969,Gibson1998}. On the other hand, degenerate parametric conversion {\it below} oscillation threshold (cavity-enhanced degenerate parametric down-conversion) is the standard approach for the generation of squeezed vacuum states of light~\cite{Wu1986,Vahlbruch2008,Vahlbruch2016,Darsow-Fromm2021,Schnabel2017}.
\section{Theoretical model}\label{results}
In order to understand the origin of laser noise suppression in DOPO at pump powers far above optical-parametric oscillation threshold, we introduce a simple model. A rigorous Hamiltonian treatment can be found in the Supplement 1.
The average output power $I_\text{out}$ of a DOPO  for $I_\text{in} \geq I_\text{th}$ is given by~\cite{breitenbach81ConversionEfficiency1995b, MMartinelli_2001}
\begin{align}
I_\text{out} = \eta I_\text{in} = 4\eta_\text{max}I_\text{th}\left(\sqrt{\frac{I_\text{in}}{I_\text{th}}}-1\right)\text{,} 
\label{Formular: Output power}
\end{align}
with threshold power $I_\text{th}$, pump power $I_\text{in}$, and conversion efficiency $\eta$, which reaches its maximum $\eta_\text{max}$ for an optimally phase-matched system, 
Next, we consider a small fluctuation $\delta I_\text{in}$ in the pump power $I_\text{in} = \overline{I_\text{in}} + \delta I_\text{in}$, where $\overline{I_\text{in}}\gg \delta I_\text{in}$ is the average power.
We also assume these fluctuations to be much stronger than quantum fluctuations (the full quantum treatment can be found in the Supplementary Information).
These fluctuations get converted into the fluctuations $\delta I_\text{out}$ of the converted power $I_\text{out} = \overline{I_\text{out}} + \delta I_\text{out}$.
Substituting this definition into \eqref{Formular: Output power}, we find the output power in terms of average value and noise
\begin{align}
\overline{I_\text{out}} + \delta I_\text{out} = 
4\eta_\text{max}I_\text{th}\left(\sqrt{\frac{\overline{I_\text{in}} + \delta I_\text{in}}{I_\text{th}}}-1\right)\text{.}
\label{Eq: Power out plus noise}
\end{align}
Since the fluctuations are much smaller than the average power, $\delta I \ll I$, we can linearize \eqref{Eq: Power out plus noise} and find the expression for the noise, keeping only the terms linear in $\delta I_{\rm in, out}$,
\begin{align}
\delta I_\text{out} = 
2\delta I_\text{in}\eta_\text{max}\sqrt{\frac{I_\text{th}}{\overline{I_\text{in}}}}\text{.}
\label{Formular: noise}
\end{align}
From \eqref{Formular: noise} we see that as the pump power grows, the converted noise $\delta I_{\rm out}$ is getting suppressed with respect to the pump noise $\delta I_{\rm in}$.
At very high input powers, the suppression of these fluctuations becomes so strong that the impact of quantum uncertainty becomes relevant.
Ultimately the noise saturates at the shot noise level, as we show in detail in the Supplement 1.
At the same time with increased noise suppression in \eqref{Formular: noise} the conversion efficiency also drops, as can be seen from \eqref{Formular: Output power}, which affects the relative level of noise.
Therefore, we consider the relative intensity noise (RIN), which is independent on the average power in the field and thus allows to compare noise levels of different systems with different powers.
\begin{equation}
{\rm{RIN}}_\text{out}\equiv
\frac{\delta I_\text{out}}{\overline{I_\text{out}}} = 
\frac{\delta I_\text{in}}{2\overline{I_\text{in}}}  \left(1-\sqrt{\frac{I_\text{th}}{\overline{I_\text{in}}}}\right)^{-1}\\
=  \rm{RIN}_\text{in}\cdot \mathcal{N}\,,
\label{Formular: Fit}
\end{equation}
where we introduced the noise transfer factor $\mathcal{N}$,
\begin{align}
\mathcal{N} = 
\frac{1}{2}
\left(
1-\sqrt{
	\frac{I_\text{th}}{\overline{I_\text{in}}}
}
\right)^{-1}.
\label{Formular: noise transfer factor}
\end{align}
This allows to define three operating regimes of an DOPO:
\begin{itemize}
	\item the \textit{weakly pumped} oscillation, $I_\text{in} \gtrapprox I_\text{th}$, where $\mathcal{N}$ increases far above 1 when approaching the threshold (see \eqref{Formular: noise transfer factor}), leading to a strong amplification of RIN on the converted field.
	\item the \textit{optimally pumped} oscillation, $I_\text{in} \approx 4 I_\text{th}$, the conversion efficiency reaches its maximum (see \eqref{Formular: Output power}) and $\mathcal{N}\approx1$: the RIN of pump and converted field are identical.
	\item the \textit{overpumped} oscillation, $I_\text{in} > 4 I_\text{th}$, where $\mathcal{N}<1$ and the RIN of the converted field is suppressed. For very high powers, $I_\text{in}\gg I_\text{th}$, $\mathcal{N} \rightarrow 0.5$, so the RIN of the converted field can be maximally suppressed by a factor of 2. Conversion efficiency gradually decreases at the same time. It should not be concluded from this result that quantum shot noise would be squeezed far above the threshold. In the supplementary material we show that far above the oscillation threshold the same applies as for any	laser: a coherent state is generated, more precisely a quasi-monochromatic field with the sideband spectrum in the vacuum state. 
\end{itemize}

\section{Experimental Testing}\label{experimental-setup}
\begin{figure}	
	\includegraphics[width=1\linewidth]{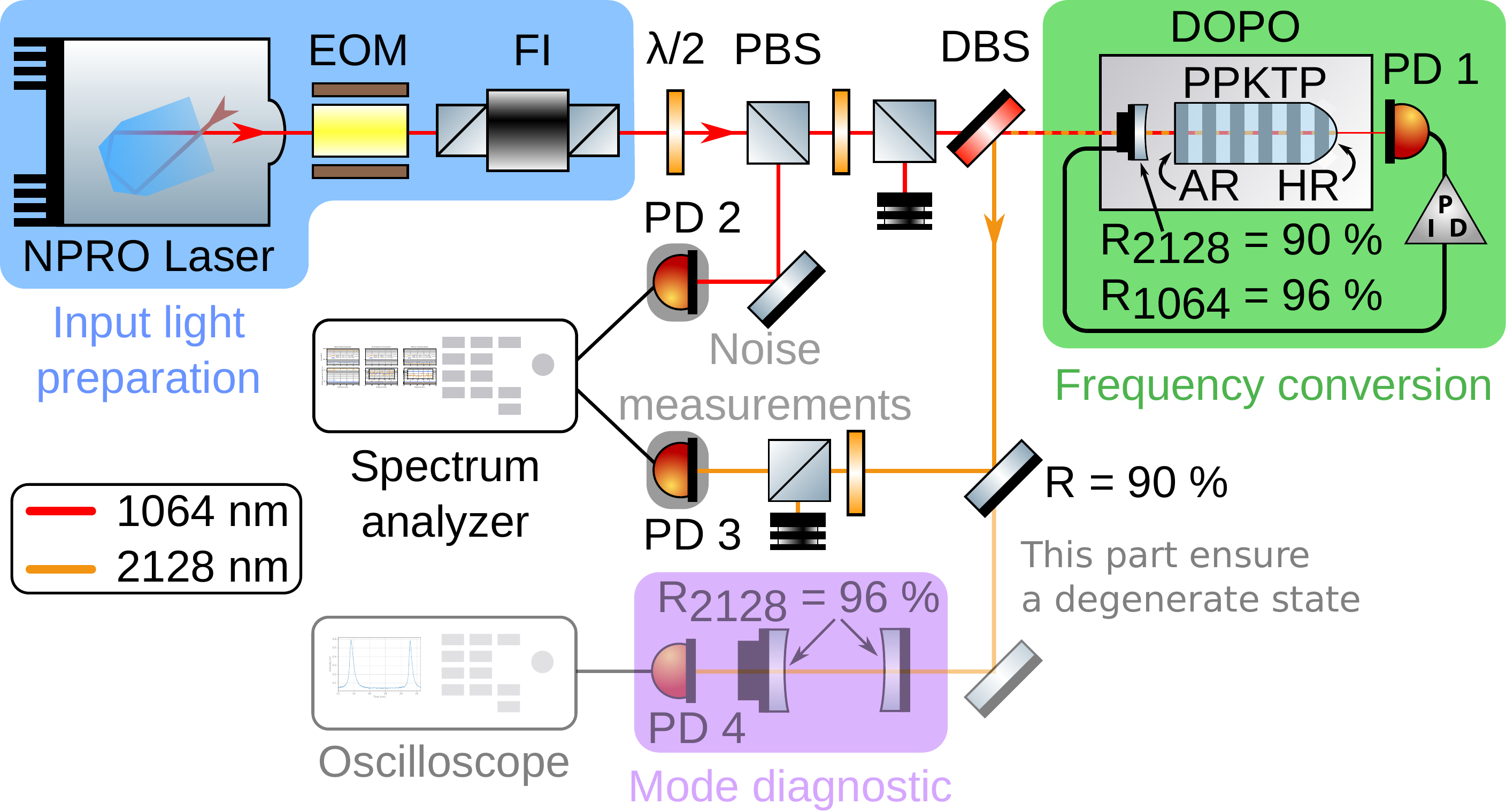}
	\caption{A \SI{1064}{\nano\metre} NPRO laser (blue box: input light preparation) pumped a degenerate optical-parametric oscillator (DOPO) to double the input wavelength to \SI{2128}{\nano\metre} (green box: frequency conversion).
		The relative intensity noise of both wavelengths was detected with amplified photo diodes (InGaAs and Ext-InGaAs, respectively) and measured with a spectrum analyzer (grey boxes: noise measurements).
		To ensure a degenerate state, the light's frequency mode content was monitored with a confocal cavity (pink box: mode diagnostic).
		NPRO: non-planar ring oscillator laser; EOM: electro-optical modulator; FI: Faraday isolator; PBS: polarizing beam-splitter; DBS: dichroic beam-splitter; DOPO: degenerate optical parametric oscillator; PD: photo detector.	
	}
	\label{fig:lasersystem}
\end{figure}
\begin{figure*}
	\centering \hspace*{-5mm}
	\includegraphics[width=0.95\linewidth]{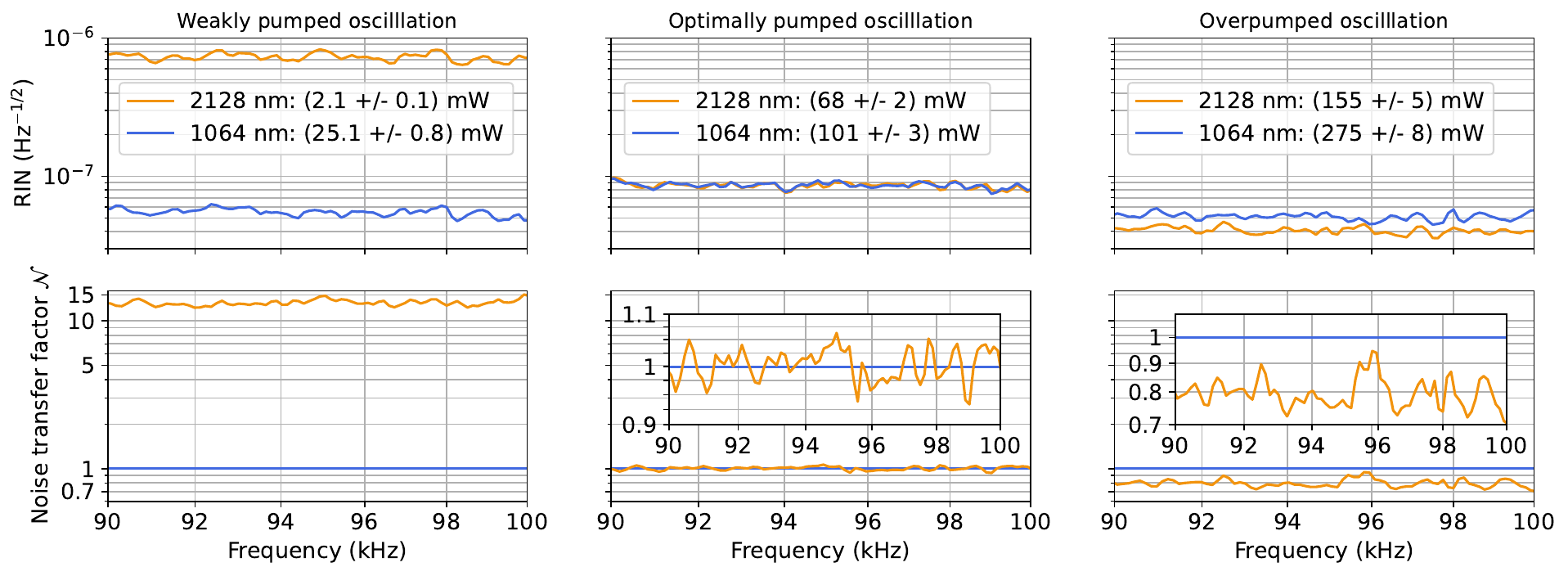}
	\caption{
		Top row: RIN measurements of the pump and converted beam for three different pump powers (left, close to threshold $I_{\text{in}}\approx \SI{25.1\pm 0.8}{\milli\watt}$; middle, at the point of highest conversion efficiency $I_{\text{in}}\approx \SI{101 \pm 3}{\milli\watt}$; right, strongly pumped $I_{\text{in}}\approx \SI{275\pm 8}{\milli\watt}$). The frequency band shown is exemplary for all frequencies well within the linewidth of the conversion resonator and was selected because it contains no peaks and lies within the optimum frequency band of the photo detector electronics.
		Bottom row: As above, showing the noise transfer factor $\mathcal{N}$ for the pump and converted light fields.
		All traces were root mean square averaged 40 times.
	}
	\label{fig:noise}
\end{figure*}
\begin{figure}
	\centering
	\includegraphics[width=1\linewidth]{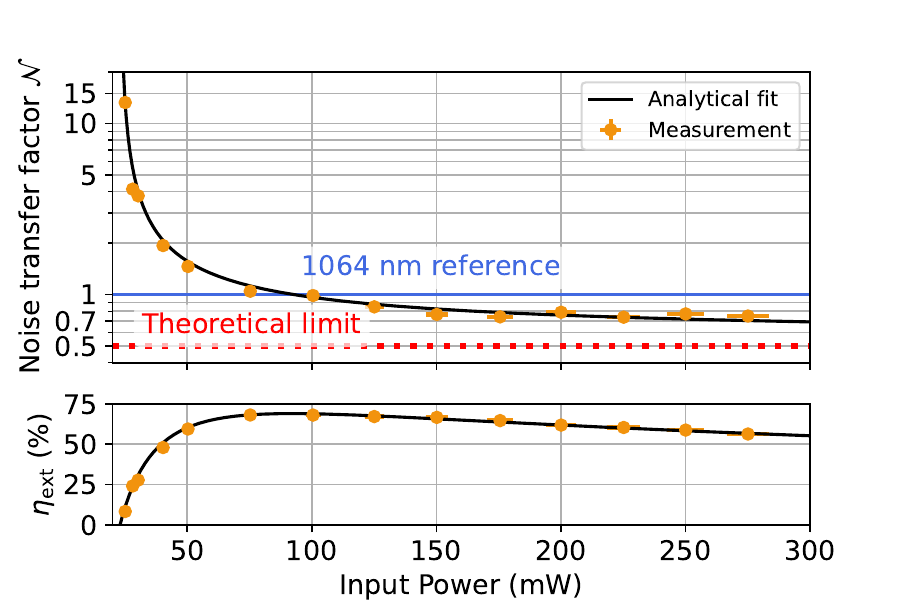}
	\caption{
		Top row:
		Relative intensity noise transfer factor $\mathcal{N}$ at different pump powers, together with the theoretical model given by \eqref{Formular: Fit}. In the limit of high pump power, the model converges to a suppression by a factor of two (red dotted line).
		The RIN of the pump field is shown as a reference (blue).
		Bottom row: External conversion efficiency ($\eta_\text{ext}$) together with the analytical fit given by \eqref{Formular: Output power}.
	}
	\label{fig:noise-suppression}
\end{figure}

We tested our theoretical predictions experimentally in a DOPO setup, shown in Figure~\ref{fig:lasersystem}.
The continuous-wave pump laser was a \SI{1064}{\nano\metre} non-planar ring oscillator (NPRO).
Its output beam passed an electro-optical modulator (EOM) for sideband generation at \SI{28}{\mega\hertz} for Pound-Drever-Hall length stabilization of the nonlinear resonator.
A Faraday isolator protected the NPRO from back reflections and back scattering.
A small amount of the pump light was extracted and measured by a photo diode (PD 2) for the RIN analysis, while the remainder was supplied to the DOPO.
The DOPO cavity featured a half-monolithic (hemilithic) design and was composed of a periodically-poled potassium titanyl phosphate (PPKTP) crystal and a separate coupling mirror, as detailed in our previous works~\cite{Darsow-Fromm2020,Darsow-Fromm2021}.
The crystal was highly-reflective coated for the pump and converted fields on its curved end face, and an anti-reflective coating for both fields on the flat front face.
The coupling mirror had reflectivities of \SI{96}\% at \SI{1064}{\nano\metre} and \SI{90}\% at \SI{2128}{\nano\metre}.
Table \ref{tab:cavity} provides the resonator parameters for both light fields.
To stabilize the length of the DOPO on resonance, we used a modified Pound-Drever-Hall control scheme in transmission (PD 1), feeding back to the piezo-mounted coupling mirror with the help of a digital control circuit~\cite{darsow-frommNQontrolOpensourcePlatform2020}.
The nonlinear parametric process of wavelength-doubling via type 0 optical-parametric amplification converted the pump field into the signal and idler light fields.
A degenerate state of the signal and idler ($\omega_\text{sig}=\omega_\text{id}$) was achieved by controlling the temperature of the nonlinear crystal as shown in~\cite{Darsow-Fromm2020} and was monitored with a linear, confocal cavity (reflectivities of \SI{96}{\percent}, PD 4).
With this setup, the DOPO's threshold was reached at a pump power of $I_\text{th} = \SI{23.6}{mW}$.
A dichroic beam-splitter separated the \SI{1064}{\nano\metre} and \SI{2128}{\nano\metre} light field after the DOPO.
The RIN of the converted light was measured with a photo diode (PD 3, extended InGaAs, Thorlabs FD05D, with custom trans\-impedance amplifier) and evaluated with a spectrum analyzer.
\begin{table}
	\centering
	\caption{Overview of the DOPO cavity parameters}
	\begin{tabular}{lrrl}
		\hline
		& \SI{1064}{\nano\metre}         &  \SI{2128}{\nano\metre} &  \\
		\hline
		waist radius  & 33.5  &  47.4   & \textmu m \\
		finesse & 153   &  59.5   & \\
		free spectral range & 3.80  &  3.83  & GHz \\
		linewidth (FWHM) & 24.9  &  64.3   & MHz \\
		coupler reflectivity & 96 & 90 & \% \\
		\hline
	\end{tabular}
	\label{tab:cavity}
\end{table}
Figure \ref{fig:noise} presents the measurement of the relative intensity noise level for different pump powers.
We show a spectrum between \SI{90}{kHz} and \SI{100}{kHz}, which was selected as it was free from spurious lines and well above photo diode dark noise for both wavelengths, however the effect could be observed across a wide spectrum corresponding to our cavity linewidth of \SI{24.9}{\mega\hertz} for \SI{1064}{\nano\metre} and \SI{64.3}{\mega\hertz} for \SI{2128}{\nano\metre}.
In the middle column, the pump power was chosen close to the point of maximum conversion efficiency, $I_\text{in} \approx 4 I_\text{th} \approx \SI{100}{mW}$.
As predicted, the RIN of pump field and converted field agreed.
For the left column, the pump power was lowered to just above the threshold power.
In this case, a significantly higher RIN of the converted field was observed, around 14 times higher than the RIN of the pump field.
Lastly, for the right column, the pump power was raised to more than 10 times the threshold power.
At these pump light levels, the RIN of the converted light fell below that of the pump light.
In our setup, we achieved a maximum noise reduction of the converted light of $\SI{25.1\pm 1.8}{\percent}$ at $\SI{275\pm 8}{\milli\watt}$ pump power.
We repeated the measurements for a wide range of pump powers, at each power obtaining the ratio between the RIN of the pump and the converted light.
The data is summarized in Figure~\ref{fig:noise-suppression} and is well fitted by the theoretical expression for $\mathcal{N}$ from \eqref{Formular: noise transfer factor}.
The error bars for the measured values was given by the \SI{3}{\percent} relative measurement error for our thermal power meter head, as specified by the manufacturer, and \SI{2}{\percent} accuracy of the homemade photo diode circuit.

\section{Conclusion}
Optical cavities are crucial components not only for advanced laser intensity stabilisation methods \cite{Kwee2012}. The strongest interferences naturally occur in the audio band and in the sub-audio frequency bands, which unfortunately fall within the linewidths of relatively easily manageable cavities. In these ranges, passive stabilisation of cavity-transmitted light is ineffective. When the relative intensity noise (RIN) marginally exceeds photon shot noise, also measurement-based active stabilisation becomes ineffective. The solution is a coherent noise suppression.\\
Optical cavities with moderate values of linewidth and finesse also play a crucial role in achieving high efficiencies in frequency conversion of for continuous-wave laser light. Our model predicts that degenerate optical parametric oscillation pumped with an optical power that is at least four times higher than the oscillation threshold power produces a cavity output field at doubled wavelength with reduced RIN of the non-quantum noise contributions compared to the short wavelength pump field. According to our model, this coherent technical noise cancellation works at any low frequency. RIN suppression is achieved without the need for light measurement, thus avoiding photon shot noise and classical sensor noise. The maximum RIN suppression is \SI{50}{\percent} and is achieved approximately in the case of infinite input power. Our experimental results corroborate our theory, demonstrating a \SI{25.1\pm 1.8}{\percent} RIN reduction at \SI{56}{\percent} external conversion efficiency producing \SI{155 \pm 5}{\milli\watt} at \SI{2128}{\nano\metre} through optical parametric wavelength doubling. We note that our theory states that the technical noise in itself is arbitrarily reduced upon wavelength doubling with an over-pumped degenerate OPO. However, since the power of the converted field also decreases in parallel, the noise quantity that is actually relevant is the RIN.\\
The transition of the popular \SI{1}{\micro\metre} wavelength of ultra-high precision laser interferometry into the \SI{2}{\micro\metre} range allows to reduce the thermal noise of cavity mirror coatings by using amorphous Si/amorphous SiN coatings on cryogenically cooled silicon mirrors and thus further improve the experimental precision \cite{steinlechnerSiliconBasedOpticalMirror2018a}. Combining our approach with wavelength doubling opens the possibility to reach the \SI{2}{\micro\metre} range by taking advantage of ultra-stable lasers in the \SI{1}{\micro\metre} range, whose RIN is already significantly reduced by state-of-the-art laser intensity stabilisation. Our approach allows to not only preserve these exceptional levels of stability, but also improve on them.

\begin{backmatter}
	\bigskip
	\noindent
This article has LIGO document number P2400033.	
	
\bmsection{Funding} This research has been funded by the Germany Federal Ministry of Education and Research, grant no. 05A20GU5.

\bmsection{Disclosures} The authors declare no conflicts of interest.

\bmsection{Data availability} Data underlying the results presented in this paper are not publicly available at this time but may be obtained from the authors upon reasonable request.

\bmsection{Supplemental document} Supporting content in the form of a detailed theoretical model can be found in Supplement 1.

\end{backmatter}

%%%%%%%%%%%%%%%%%%%%%%%%%%%%%%%%%%%%%%% SUPPLEMENTARY
%%%%%%%%%%%%%%%%%%%%%%%%%%%%%%%%%%%%%%% SUPPLEMENTARY
%%%%%%%%%%%%%%%%%%%%%%%%%%%%%%%%%%%%%%% SUPPLEMENTARY
%%%%%%%%%%%%%%%%%%%%%%%%%%%%%%%%%%%%%%% SUPPLEMENTARY

\newcommand{\beginsupplement}{%
	\setcounter{table}{0}
	\renewcommand{\thetable}{S\arabic{table}}%
	\setcounter{figure}{0}
	\renewcommand{\thefigure}{S\arabic{figure}}%
	\renewcommand{\theequation}{S\arabic{equation}}
	\setcounter{equation}{0}
	\renewcommand{\thesection}{S\@arabic{section}}
	\setcounter{equation}{0}
}

\section{Supplementary Material}
\beginsupplement
In this Supplement, we calculate the conversion efficiency, noise contributions and the noise transfer factor for the RIN, according to \eqref{Formular: noise transfer factor}.
We start with the Hamiltonian for a cavity with a nonlinear crystal and two intra-cavity fields signal $\hat{s}$ and pump $\hat{p}$~\cite{drummond2004quantum} for perfectly matching input and cavitiy modes and zero loss:
\begin{align}
\mathcal{H} = &\hbar \omega_s \hat{s}^\dag \hat{s}
+ \hbar \omega_p \hat{p}^\dag \hat{p}
+ i \hbar \chi \left(\hat{p}^\dag \hat{s}^2 -h.c.\right) + \nonumber\\
& i \hbar \sqrt{2\gamma_s} \int_{-\infty}^{\infty}\left(\hat{s}^\dag(\omega) \hat{s}_{\rm in}(\omega)-h.c.\right)d\omega +\\
& i \hbar \sqrt{2\gamma_p} \int_{-\infty}^{\infty}\left(\hat{p}^\dag(\omega) \hat{p}_{\rm in}(\omega)-h.c.\right)d\omega \nonumber\text{,}
\end{align}
where $\omega_{s,p}$ is the angular frequency of the respective light field, $\gamma_{s,p}$ is the coupling rate of the cavity for the corresponding light field $\hat{s}_{\rm in},\,\hat{p}_{\rm in}$, and $\chi$ is the second-order nonlinear coefficient of the crystal.
The Heisenberg equations of motion are given by
\begin{align}\label{eom}
\dot{\hat{s}} &= -\gamma_s \hat{s} + \sqrt{2\gamma_s} \hat{s}_{\rm in} - 2 \chi \hat{s}^\dag \hat{p},\\
\dot{\hat{p}} &= -\gamma_p \hat{p}+ \sqrt{2\gamma_p} \hat{p}_{\rm in} + \chi \hat{s}^2, \\
\hat{s}_{\rm out} &= -\hat{s}_{\rm in} + \sqrt{2\gamma_s}\hat{s}\text{.}
\end{align}
We can linearize these equations, considering small fluctuations around the average amplitude: $\hat{s} = S + \hat{\varsigma}$,\, $\hat{p} = P + \hat{\rho}$,\, $\hat{\varsigma}\ll S, \,\hat{\rho}\ll P$.
\subsection{Average power}
We find the average power considering the steady-state approximation, which is possible in our application, where there are no instabilities and chaotic behavior
\begin{align}
0 &= -\gamma_s S + \sqrt{2\gamma_s} S_{\rm in} - 2\chi S^* P,\\
0 &= -\gamma_pP + \sqrt{2\gamma_p} P_{\rm in} + \chi  S^2,\\
S_{\rm out} &= -S_{\rm in} + \sqrt{2\gamma_s}S\text{.}
\end{align}
We solve equations for $P$ and substitute it into equation for $S$:
\begin{align}
P &= \sqrt{\frac{2}{\gamma_p}}P_{\rm in}+\frac{\chi}{\gamma_p}S^2,\\
0 &= -\gamma_s S
+\sqrt{2\gamma_s}S_{\rm in}
-2\chi\sqrt{\frac{2}{\gamma_p}}S^* P_{\rm in}
- \frac{(\chi)^2}{\gamma_p} S^* S^2\text{.}
\label{eq:noise}
\end{align}
Applied to our experiment, where there is no input field $S_{\rm in}=0$, we can define the average signal power:
\begin{equation}
I_s = \frac{\hbar \omega_s}{2} S^* S = \frac{\hbar \omega_s\gamma_p}{(\chi)^2}\left(-\gamma_s-2\chi\sqrt{\frac{2}{\gamma_p}}\frac{S^*}{S}P_{\rm in}\right)\text{.}
\end{equation}
We choose the phases of the fields in such a way that the power is the positive value. We choose $P_{\rm in}$ to be real, and $S=|S|e^{i\phi}$, where $\phi=\pi/2$.
With this we can find the condition for the threshold amplitude $P_{\rm th}$ and power $I_{\rm th}$, where $I_s = 0$:
\begin{align}
P_{\rm th} = \frac{\gamma_s\sqrt{\gamma_p}}{2\sqrt{2}\chi}, \quad I_{\rm th} = \frac{\hbar \omega_s\gamma_s^2\gamma_p}{8\chi^2}
\end{align}
Then equation for the output signal power (using that $S_{\rm out} = \sqrt{2\gamma_s}S$) can be written as:
\begin{equation}\label{power}
I_{\rm s,out} = 2\hbar\omega_sP_{\rm th}^2 \left(\frac{P_{\rm in}}{P_{\rm th}}-1\right)
=4I_{\rm th}\left(\sqrt{\frac{I_{\rm in}}{I_{\rm th}}}-1\right).
\end{equation}
Conversion efficiency is defined as the ratio of the two powers:
\begin{equation}
\eta = \frac{I_{\rm s,out}}{I_{\rm in}} = 4 \frac{I_{\rm th}}{I_{\rm in}}\left(\sqrt{\frac{I_{\rm in}}{I_{\rm th}}}-1\right).
\end{equation}
It reaches its maximum for $I_{\rm in}=4I_{\rm th}$. In the idealized situation without losses and with perfect phase matching considered here, conversion efficiency reaches unity.

\subsection{Noise}
In order to compute the noise behavior in the system, we linearize equations of motion, keeping only the terms of linear in noise:
\begin{align}
\hat{\rho} =&\sqrt{\frac{2}{\gamma_p}}\hat{\rho}_{\rm in} - \frac{2\chi}{\gamma_p} S \hat{\varsigma},\\
\begin{split}
0=&- \frac{2 \chi^2}{\gamma_b} S^2 \left(\hat{\varsigma}^\dagger + 2 \hat{\varsigma}\right) +  2 \chi \sqrt{\frac{2}{\gamma_b}} \left(S \hat{\rho} + P \hat{\varsigma}^\dagger\right) \\&- \gamma_a \hat{\varsigma} + \sqrt{2 \gamma_{a}} \hat{\varsigma}_{\text{in}}.
\end{split}
\end{align}
We define amplitude and phase quadratures amplitudes:
\begin{equation}
\hat{\varsigma}_a = \frac{\hat{\varsigma} + \hat{\varsigma}^\dag}{\sqrt{2}},\quad \hat{\varsigma}_p = \frac{\hat{\varsigma} - \hat{\varsigma}^\dag}{i\sqrt{2}},
\end{equation}
where the factor $ \sqrt{2}$ normalizes the respective ground state variances of a one-sided spectrum to unity. The solutions for the intra-cavity fields read
\begin{align}
&\hat{\varsigma}_a = \hat{\varsigma}_{\rm in,a}\frac{P_{\rm th}}{\sqrt{2\gamma_p}\left(P_{\rm in} - P_{\rm th}\right)} + \hat{\rho}_{\rm in,a}\frac{\sqrt{P_{\rm th}}}{\sqrt{\gamma_s\left(P_{\rm in}-P_{\rm th}\right)}}\\
&\hat{\varsigma}_p = \hat{\varsigma}_{\rm in,p}\frac{P_{\rm th}}{\sqrt{2\gamma_p}P_{\rm in}} + \hat{\rho}_{\rm in,p}\frac{\sqrt{P_{\rm th}\left(P_{\rm in}-P_{\rm th}\right)}}{\sqrt{\gamma_s}P_{\rm in}}.
\end{align}
Now we use input-output relations and compute the noise at the output of the cavity:
\begin{align}
&\hat{\varsigma}_{\rm out,a} = -\hat{\varsigma}_{\rm in,a}\frac{P_{\rm in}-2P_{\rm th}}{P_{\rm in} - P_{\rm th}} + \hat{\rho}_{\rm in,a}\frac{\sqrt{2P_{\rm th}}}{\sqrt{\left(P_{\rm in}-P_{\rm th}\right)}}\\
&\hat{\varsigma}_{\rm out,p} = -\hat{\varsigma}_{\rm in,p}\frac{P_{\rm in}-P_{\rm th}}{P_{\rm in}} + \hat{\rho}_{\rm in,p}\frac{\sqrt{2P_{\rm th}\left(P_{\rm in}-P_{\rm th}\right)}}{P_{\rm in}}.
\end{align}
We can finally compute the single-sided spectral densities of the noises at Fourier frequencies close to zero:
\begin{align}\label{variances}
&\mathcal{S}_{\hat{\varsigma}, \rm{out}}^{\rm a} = \mathcal{S}_{\hat{\varsigma}, \rm{in}}^{\rm a}\frac{\left(P_{\rm in}-2P_{\rm th}\right)^2}{\left(P_{\rm in} - P_{\rm th}\right)^2} + \mathcal{S}_{\hat{\rho}, \rm{in}}^{\rm a}\frac{2P_{\rm th}}{\left(P_{\rm in}-P_{\rm th}\right)},\\
&\mathcal{S}_{\hat{\varsigma}, \rm{out}}^{\rm p} = \mathcal{S}_{\hat{\varsigma}, \rm{in}}^{\rm p}\frac{\left(P_{\rm in}-P_{\rm th}\right)^2}{P_{\rm in}^2} + \mathcal{S}_{\hat{\rho}, \rm{in}}^{\rm p}\frac{2P_{\rm th}\left(P_{\rm in}-P_{\rm th}\right)}{P_{\rm in}^2}.
\end{align}

We now consider the relevant case where the signal input is in the ground state, and the pump beam has both quantum and technical (and thermal) excitations:
\begin{align}
	\mathcal{S}_{\hat{\varsigma}, \rm{in}}^{\rm a} = \mathcal{S}_{\hat{\varsigma}, \rm{in}}^{\rm p} = 1,\\
	\mathcal{S}_{\hat{\rho}, \rm{in}}^{\rm a} = 1 + \mathcal{S}_{\hat{\rho}, \rm{tech}}^{\rm a},\\
	\mathcal{S}_{\hat{\rho}, \rm{in}}^{\rm p} = 1 + \mathcal{S}_{\hat{\rho}, \rm{tech}}^{\rm p},
\end{align}
where $\mathcal{S}_{\hat{\rho}, \rm{in}}^{\rm a},\, \mathcal{S}_{\hat{\rho}, \rm{in}}^{\rm p}$ are the spectral densities of amplitude and phase technical noise.
With this we can re-write Eq.\,\ref{variances} in a simpler form:   
\begin{align}
&\mathcal{S}_{\hat{\varsigma}, \rm{out}}^{\rm a} = 1 + \frac{P_{\rm th}^2}{\left(P_{\rm in} - P_{\rm th}\right)^2} + \frac{2P_{\rm th}}{\left(P_{\rm in}-P_{\rm th}\right)}\mathcal{S}_{\hat{\rho}, \rm{tech}}^{\rm a},\\
&\mathcal{S}_{\hat{\varsigma}, \rm{out}}^{\rm p} = 1 - \frac{P_{\rm th}^2}{P_{\rm in}^2} + \frac{2P_{\rm th}\left(P_{\rm in}-P_{\rm th}\right)}{P_{\rm in}^2}\mathcal{S}_{\hat{\rho}, \rm{tech}}^{\rm p}.
\end{align}
In the first two terms of these equations we see the well-known strong anti-squeezing in amplitude and squeezing in phase for pump powers close to threshold, which quickly approaches shot noise level far above threshold.
At the same time, we see that the pump amplitude noise $\mathcal{S}_{\hat{\rho}, \rm{tech}}^{\rm a}$ gets amplified close to threshold, and heavily suppressed far above threshold.
Ultimately, regardless of the magnitude classical contribution $\mathcal{S}_{\hat{\rho}, \rm{tech}}^{\rm a}, \mathcal{S}_{\hat{\rho}, \rm{tech}}^{\rm p}$, it will get suppressed down to the shot noise level. Far above the threshold, the output light approaches the coherent state. Squeezing does not happen.

In order to compute RIN, we define the noise $\hat{n}_{\rm s,out}$ on the output power of the signal $I_{\rm s,out}$:
\begin{equation}
\hat{n}_{\rm s,out} = \hbar \omega_s S_{\rm out} \hat{\varsigma}_{\rm out,a} = \sqrt{2 \hbar \omega_s I_{\rm s,out}}\hat{\varsigma}_{\rm out,a},
\end{equation}
from which we obtain the RIN:
\begin{equation}
{\rm RIN}_{\rm out} = \frac{n_{\rm s,out}}{I_{\rm s,out}} = \frac{\sqrt{2 \hbar \omega_s}}{\sqrt{I_{\rm s,out}}} \hat{\varsigma}_{\rm out,a}.
\end{equation}
In the following discussion we limit ourselves to the case of the large technical noise $\mathcal{S}_{\hat{\rho}, \rm{tech}}^{\rm a}\gg 1$, so we ignore the contribution of quantum noise. 
This allows to compute the spectral density of the noise $\hat{n}_{\rm s,out}$:

\begin{equation}
\begin{aligned}
\frac{\mathcal{S}_{n, \rm{out}}}{I_{\rm s,out}^2} &= \frac{{2 \hbar \omega_s}}{{I_{\rm s,out}}} \mathcal{S}_{\hat{\varsigma}, \rm{out}}^{\rm a}
\\&= \frac{{2 \hbar \omega_s}}{{I_{\rm s,out}}} {\frac{2P_{\rm th}}{\left(P_{\rm in}-P_{\rm th}\right)}}\mathcal{S}_{\hat{\rho}, \rm{tech}}^{\rm a} 
\\&= \frac{{2 \hbar \omega_s}}{{I_{\rm s,out}}} {\frac{2\sqrt{I_{\rm th}}}{\left(\sqrt{I_{\rm in}}-\sqrt{I_{\rm th}}\right)}}\frac{\mathcal{S}_{n, \rm{in}}}{{2\hbar \omega_p I_{\rm in}}}
\\ &= {\frac{I_{\rm in}}{I_{\rm s, out}}\frac{\sqrt{I_{\rm th}}}{\sqrt{I_{\rm in}}-\sqrt{I_{\rm th}}}}\frac{\mathcal{S}_{n, \rm{in}}}{I_{\rm in}^2},
\end{aligned}
\end{equation}
where we defined the spectral density of the input technical noise on laser power $\mathcal{S}_{n, \rm{in}} = 2\hbar \omega_p I_{\rm in}\mathcal{S}_{\hat{\rho}, \rm{tech}}^{\rm a}$.
We can define the noise transfer factor $\mathcal{N}$:
\begin{equation}
	\frac{\mathcal{S}_{n, \rm{out}}}{I_{\rm s,out}^2} = \mathcal{N}^2 \frac{\mathcal{S}_{n, \rm{in}}}{I_{\rm in}^2}
\end{equation}
From this definition and \eqref{power} we arrive at noise transfer factor (\eqref{Formular: noise transfer factor} in the main text):
\begin{equation}
\mathcal{N} = \sqrt{\frac{I_{\rm in}}{I_{s, out}}\frac{\sqrt{I_{\rm th}}}{\sqrt{I_{\rm in}}-\sqrt{I_{\rm th}}}} = \frac{1}{2}\frac{\sqrt{I_{\rm in}}}{\sqrt{I_{\rm in}}-\sqrt{I_{\rm th}}}.
\end{equation}
The degenerate optical parametric oscillator, which is pumped with infinite power ("overpumped"), generates a field (with twice the wavelength) with a halved RIN value.
Importantly, while the noise itself is suppressed infinitely, the RIN is not, with a limit of 2 on suppression.

\bibliography{references}
\appendix

\end{document}